\documentclass[english,preprint, aps]{revtex4}
\usepackage{lmodern}
\usepackage[T1]{fontenc}
\usepackage[latin9]{inputenc}
\usepackage{color}
\usepackage{graphicx}
\usepackage{epstopdf}

\makeatletter

\@ifundefined{textcolor}{}
{%
 \definecolor{BLACK}{gray}{0}
 \definecolor{WHITE}{gray}{1}
 \definecolor{RED}{rgb}{1,0,0}
 \definecolor{GREEN}{rgb}{0,1,0}
 \definecolor{BLUE}{rgb}{0,0,1}
 \definecolor{CYAN}{cmyk}{1,0,0,0}
 \definecolor{MAGENTA}{cmyk}{0,1,0,0}
 \definecolor{YELLOW}{cmyk}{0,0,1,0}
 }

\makeatother

\usepackage{babel}

\begin{document}

\title{Onset of Fast Reconnection in Hall Magnetohydrodynamics Mediated
by the Plasmoid Instability}

\author{Yi-Min Huang }

\email{yimin.huang@unh.edu}

\author{A. Bhattacharjee}

\author{Brian P. Sullivan}

\affiliation{Center for Integrated Computation and Analysis of Reconnection and
Turbulence and Center for Magnetic Self-Organization in Laboratory
and Astrophysical Plasmas, University of New Hampshire, Durham, NH
03824}
\begin{abstract}
The role of a super-Alfv\'enic plasmoid instability in the onset
of fast reconnection is studied by means of the largest Hall magnetohydrodynamics
simulations to date, with system sizes up to $10^{4}$ ion skin depths
($d_{i}$). It is demonstrated that the plasmoid instability can facilitate
the onset of rapid Hall reconnection, in a regime where the onset
would otherwise be inaccessible because the Sweet-Parker width is
significantly above $d_{i}$. However, the topology of Hall reconnection
is not inevitably a single stable X-point. There exists an intermediate
regime where the single X-point topology itself exhibits instability,
causing the system to alternate between a single X-point geometry
and an extended current sheet with multiple X-points produced by the
plasmoid instability. Through a series of simulations with various
system sizes relative to $d_{i}$, it is shown that system size affects
the accessibility of the intermediate regime. The larger the system
size is, the easier it is to realize the intermediate regime. Although
our Hall MHD model lacks many important physical effects included
in fully kinetic models, the fact that a single X-point geometry is
not inevitable raises the interesting possibility for the first time
that Hall MHD simulations may have the potential to realize reconnection
with geometrical features similar to those seen in fully kinetic simulations,
namely, extended current sheets and plasmoid formation. 
\end{abstract}
\maketitle

\section{Introduction}

Magnetic reconnection is thought to be the underlying mechanism that
powers explosive events such as flares, substorms, and sawtooth crashes
in fusion plasmas\cite{ZweibelY2009}\textbf{.} Such events commonly
feature impulsive onset, whereby the system exhibits a sudden increase
in the reconnection rate\cite{BhattacharjeeMW2001}. In classic Sweet-Parker
theory\cite{Sweet1958,Parker1963}, based on resistive magnetohydrodynamics
(MHD), the reconnection site has the structure of a thin current sheet
of length $L$, which is of the order of the system size, and a width
$\delta_{SP}\simeq L/\sqrt{S}$, where the Lundquist number $S$ is
related to the length $L$, the Alfv\'en speed $V_{A}$, and the
resistivity $\eta$ by the relation $S\equiv LV_{A}/\eta$.  The plasma
outflow speed from the reconnection site is approximately $V_{A}$,
and the inflow speed, which is a measure of the reconnection rate,
is approximately $V_{A}/S^{1/2}$ under quasi-steady conditions. In
most plasmas of interest, the Lundquist numbers are very high. Consequently,
the Sweet-Parker reconnection rates are usually several orders of
magnitude too slow to account for the observed rate of energy release
after onset. The strong dependence of the reconnection rate on $S$
in the Sweet-Parker theory has led to a broad consensus that the solution
to the onset problem for high-$S$ plasmas lies outside the domain
of resistive MHD, and requires the inclusion of collisionless effects.
In particular, for two-dimensional (2D) configurations without a guide
field, a precise criterion has been proposed that accounts for a slow
growth phase (identified as a Sweet-Parker phase in many cases of
interest), followed by rapid onset caused by the Hall current, which
is a signature of the decoupling of electron and ion motion\textcolor{black}{{}
at scales below the ion skin depth $d_{i}$ \cite{MaB1996a,DorelliB2003,Bhattacharjee2004,CassakSD2005}.
}(Here $d_{i}=c/\omega_{pi}$, where $c$ is the speed of light and
$\omega_{pi}$ is the ion plasma frequency.) The criterion predicts
that when $\delta_{SP}<d_{i}$, the system will spontaneously make
a transition to a rapid reconnection phase, with an inflow velocity
$\sim0.1V_{A}$. This criterion has been tested extensively by numerical
simulations\textcolor{black}{\cite{MaB1996a,DorelliB2003,Bhattacharjee2004,CassakSD2005}}
as well as controlled laboratory experiments\cite{YamadaRJBGKK2006}.

The recent discovery of a linear, super-Alfv\'enic plasmoid instability\cite{LoureiroSC2007}
in high-$S$ plasmas raises qualitatively new questions for the criterion
stated above. It has long been known that the Sweet-Parker reconnection
layer can become unstable to a secondary tearing instability. However,
only recently has a precise linear study shown that the linear growth
rate $\gamma$ of the instability scales as $\gamma\sim S^{1/4}(V_{A}/L)$.
The positive exponent of $S$ yields high growth rates for high $S$
plasmas, whereas most resistive instabilities scale with $S$ to some
negative fractional power. This seemingly counterintuitive result
can actually be deduced from the dispersion relation for classical
tearing modes\cite{CoppiGPRR1976} with one crucial new insight: the
Sweet-Parker layer supports an increasingly singular current sheet
as $S\to\infty$.\cite{BhattacharjeeHYR2009} Furthermore, even within
the framework of resistive MHD, this linear instability leads to a
nonlinear regime where the reconnection rate becomes nearly independent
of $S$, with an inflow velocity $\sim10^{-2}V_{A}$.\cite{BhattacharjeeHYR2009,HuangB2010}
The original Sweet-Parker current sheet breaks up into a chain of
plasmoids and a sequence of shorter but thinner current sheets, with
widths much smaller than $\delta_{SP}$.\cite{BhattacharjeeHYR2009,DaughtonRAKYB2009,CassakSD2009} 

The presence of the plasmoid instability in high-$S$ systems uncovers
a deep flaw in the Sweet-Parker model, and raises questions about
the conventional scenario of the onset of Hall reconnection. Because
secondary current sheets are thinner than the primary Sweet-Parker
current sheet, potentially they may trigger onset of Hall MHD (or
kinetic) reconnection when the widths reach the $d_{i}$ scale, even
in systems where the original onset criterion $\delta_{SP}<d_{i}$
is not met. Shibata and Tanuma \cite{ShibataT2001}  proposed just
this scenario in an insightful paper, years before the recent spate
of interest in this topic. Recently, numerical studies have been carried
out with fully kinetic particle-in-cell (PIC) simulations including
a collision operator by Daughton \emph{et al.} \cite{DaughtonRAKYB2009}
and Hall MHD simulations by Shepherd and Cassak \cite{ShepherdC2010}
confirming the role of the plamoid instability in triggering onset
of Hall (or kinetic) reconnection. However, the PIC and Hall MHD simulations,
discussed in Refs. \cite{DaughtonRAKYB2009} and \cite{ShepherdC2010},
show qualitatively very different behaviors after onset. Whereas PIC
simulations continue to exhibit copious generation of plasmoids, the
Hall MHD solutions appear to settle down to a single stable X-point
state with all plasmoids expelled. This apparent qualitative difference
between the two types of simulations raises the following important
questions of principle: Does the onset of Hall reconnection in Hall
MHD models inevitably lead to a Hall current dominated regime in which
all plasmoids are expelled? Can Hall MHD realize current sheet geometries
qualitatively similar to those seen in fully kinetic simulations,
where new plasmoids are constantly generated? 

In this paper, we address these questions by means of the largest
two-dimensional resistive Hall MHD reconnection simulations ever carried
out, with the ratio $L/d_{i}$ ranging from $2.5\times10^{3}$ to
$1.0\times10^{4}$, in a configuration of two coalescing magnetic
islands. We confirm the previous results that the plasmoid instability
can trigger the onset of Hall MHD reconnection in systems that do
not meet the criterion $\delta_{SP}<d_{i}$ for onset. In addition,
we demonstrate that the topology of Hall MHD reconnection is not inevitably
a single stable X-point. There exists an intermediate regime where
the single X-point topology itself exhibits instability, causing the
system to alternate between a single X-point and an extended current
sheet with multiple X-points produced by the plasmoid instability.
Furthermore, through a series of simulations with various system sizes
relative to $d_{i}$, we show numerical evidence supporting the idea
that system size affects the accessibility of the intermediate regime.
The larger the system size is, the easier it is to realize the intermediate
regime.

The present study employs a simple Hall MHD model, which clearly lacks
many important physical effects that are included in fully kinetic
models. A constant resistivity instead of the Spitzer resistivity\cite{Braginskii1965}
is employed; an isothermal equation of state is employed and Ohmic
heating is neglected; plasma pressure is assumed to be a scalar rather
than a tensor; and electrons are assumed to be massless. Furthermore,
it has been argued that in many collisionless or weakly collisional
systems of interest the reconnection electric field typically exceeds
the Dreicer runaway field,\cite{Dreicer1959} therefore classical
resistivity cannot play a significant role. While these limitations
merit further investigations, the fact that a single X-point geometry
is not inevitable in Hall MHD simulations raises the interesting possibility
for the first time that Hall MHD simulations may have the potential
to realize reconnection with geometrical features similar to those
seen in fully kinetic simulations, namely, extended current sheets
and plasmoid formation.

\section{Numerical Model}

Our simulations are based on resistive Hall MHD equations. These equations
in normalized form are: \begin{equation}
\partial_{t}\rho+\nabla\cdot\left(\rho\mathbf{u}\right)=0,\label{eq:2}\end{equation}
\begin{equation}
\partial_{t}(\rho\mathbf{u})+\nabla\cdot\left(\rho\mathbf{uu}\right)=-\nabla p+\mathbf{J}\times\mathbf{B}+\epsilon\mathbf{f}(\mathbf{x},t),\label{eq:3}\end{equation}
\begin{equation}
\partial_{t}\mathbf{B}=-\nabla\times\mathbf{E},\label{eq:4}\end{equation}
\begin{equation}
\mathbf{u}_{e}=\mathbf{u}-d_{i}\frac{\mathbf{J}}{\rho},\label{eq:ve}\end{equation}
\begin{equation}
\mathbf{E}=-\mathbf{u}_{e}\times\mathbf{B}-d_{i}\frac{\nabla p_{e}}{\rho}+\eta\mathbf{J},\label{eq:Ohm}\end{equation}
where $\rho$\textcolor{black}{{} is the plasma density, $\mathbf{u}$
is the ion velocity, $\mathbf{u}_{e}$ is the electron velocity, $p$
is the total pressure, $p_{e}$ is the electron pressure, $\mathbf{B}$
is the magnetic field, $\mathbf{E}$ is the electric field, $\mathbf{J}=\nabla\times\mathbf{B}$
is the electric current density, $\eta$ is the resistivity, and $d_{i}$
is the ion skin depth. Isothermal equations of state are assumed,
}\textcolor{black}{\emph{i.e.}}\textcolor{black}{{} $p_{e}=p_{i}=\rho T$,
where $p_{i}$ is ion pressure, and $T$ is a constant temperature.
The total pressure is $p=p_{e}+p_{i}=2\rho T$. Electron inertia terms
are neglected in the generalized Ohm's law, Eq. (\ref{eq:Ohm}). The
electron pressure term $-d_{i}\nabla p_{e}/\rho$ has been omitted
in this study, because it does not contribute to the dynamics after
taking the curl of $\mathbf{E}$ in Eq. (\ref{eq:4}), due to the
isothermal equation of state. A weak random forcing term $\epsilon\mathbf{f}$
is added to the ion momentum equation, as was done in a previous study.\cite{HuangB2010}
The normalizations of Eqs. (\ref{eq:2}) -- (\ref{eq:Ohm}) are based
on constant reference values of the density $n_{0}$, and the magnetic
field $B_{0}$. Lengths are normalized to the system size $L$, and
time is normalized to the global Alfv\'en time $t_{A}=L/V_{A}$,
where $V_{A}=B_{0}/\sqrt{4\pi n_{0}m_{i}}$ and $m_{i}$ is the ion
mass. The normalizations of physical variables are given by (normalized
$\to$ physical units): $\rho\to\rho/n_{0}m_{i}$, $\mathbf{B}\to\mathbf{B}/B_{0}$,
}\textbf{\textcolor{black}{$\mathbf{E}\to c\mathbf{E}/B_{0}V_{A}$}}\textcolor{black}{,
$\mathbf{u}\to\mathbf{u}/V_{A}$, $p\to p/n_{0}m_{i}V_{A}^{2}$, $\mathbf{J}\to\mathbf{J}/(B_{0}c/4\pi L)$,
and $d_{i}\to d_{i}/L\equiv\sqrt{m_{i}c^{2}/4\pi n_{0}e^{2}}/L$.
In 2D simulations, the magnetic field is expressed in terms of the
flux function $\psi$ and the out-of-plane component $B_{y}$ as $\mathbf{B}=\nabla\psi\times\mathbf{\hat{y}}+B_{y}\mathbf{\hat{y}}$.
The variables $\psi$ and $B_{y}$ are stepped in the code. The governing
equations are numerically solved with a massively parallel code HMHD,
which is a two-fluid extension of the resistive MHD code used in previous
studies.\cite{BhattacharjeeHYR2009,HuangB2010} }The numerical algorithm
\cite{GuzdarDMHL1993} approximates spatial derivatives by finite
differences with a five-point stencil in each direction. The time-stepping
scheme can be chosen from several options including a second-order
accurate trapezoidal leapfrog method and various strong stability
preserving Runge--Kutta methods.\cite{GottliebST2001,SpiteriR2002}
We employ the second-order accurate trapezoidal leapfrog method in
this study. HMHD has the capability of nonuniform meshes that allows
better resolution of the reconnection layer.

We employ the same simulation setup of two coalescing magnetic islands
as in a previous study\cite{HuangB2010}.\textcolor{black}{{} }The 2D
simulation box is \textcolor{black}{the domain $(x,z)\in[-1/2,1/2]\times[-1/2,1/2]$}.
In normalized units, the initial magnetic field \textcolor{black}{is
given by $\mathbf{B}_{0}=\nabla\psi_{0}\times\mathbf{\hat{y}}$, where
}$\psi_{0}=\tanh\left(z/h\right)\cos\left(\pi x\right)\sin\left(2\pi z\right)/2\pi$.
The parameter $h$, which is set to $0.01$ for all simulations, determines
the initial current layer width. The initial plasma density $\rho$
is approximately $1$, and the plasma temperature $T$ is $3$. The
density profile has a weak nonuniformity such that the initial condition
is approximately force-balanced. The initial peak magnetic field and
Alfv\'en speed are both approximately unity. The plasma beta $\beta\equiv p/B^{2}=2\rho T/B^{2}$
is greater than $6$ everywhere. Perfectly conducting and free slipping
boundary conditions are imposed along both $x$ and $z$ directions.
Specifically, \textcolor{black}{we have $\psi=0$, $\mathbf{u}\cdot\mathbf{\hat{n}}=0$,
$\mathbf{\hat{n}}\cdot\nabla\left(\mathbf{\hat{n}}\times\mathbf{u}\right)=0$,
$\mathbf{\hat{n}}\cdot\nabla\rho=0$, and $B_{y}=0$ on the boundaries
(here $\mathbf{\hat{n}}$ is the unit normal vector to the boundary).
Only the upper half of the domain ($z\ge0$) is simulated, and solutions
in the lower half are inferred by symmetries. The computational mesh
consists of $6400\times1024$ grid points. The grid points along $z$
are strongly concentrated around $z=0$, with the smallest grid size
$\Delta z=1.4\times10^{-5}.$ The grid points along $x$ are weakly
nonuniform, with the smallest grid size $\Delta x=1.2\times10^{-4}$
at $x=0$. For this system, the critical Lundquist number $S_{c}$
for onset of the plasmoid instability is approximately $4\times10^{4}$
in resistive MHD} ($d_{i}=0$)\textcolor{black}{.}\cite{HuangB2010}\textcolor{black}{{} }

\section{A Phase Diagram}

\begin{figure}
\begin{centering}
\includegraphics[scale=0.6]{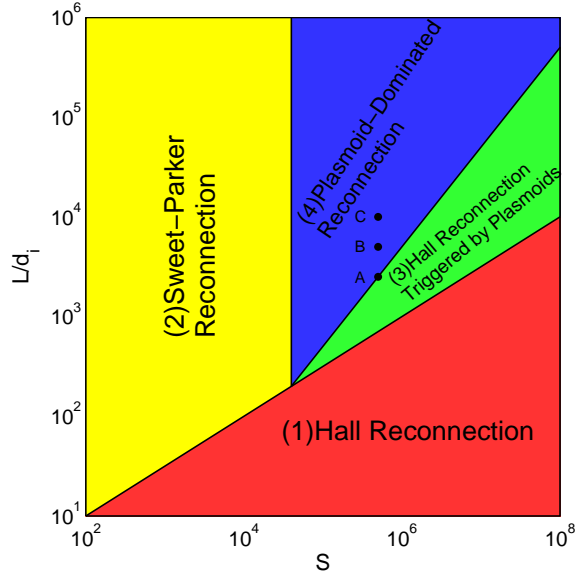}
\par\end{centering}

\caption{(Color online) The parameter space in the phase diagram is divided
into four regions. \textcolor{black}{(1) Hall reconnection: $d_{i}>\delta_{SP}$.
(2) Sweet-Parker reconnection: $d_{i}<\delta_{SP}$ and $S<S_{c}$.
(3) Hall reconnection triggered by plasmoids: $\delta_{SP}(S_{c}/S)^{1/2}<d_{i}<\delta_{SP}$.
(4) Plasmoid-dominated reconnection: $S>S_{c}$ and $\delta_{SP}(S_{c}/S)^{1/2}>d_{i}$.}
The dots denote the parameters for three different runs. All three
runs have $S=5\times10^{5}$. The parameter $L/d_{i}$ is $2.5\times10^{3}$
for Run A, $5\times10^{3}$ for Run B, and $10^{4}$ for Run C, respectively.
A fourth run, Run D, from a previous resistive MHD study\cite{HuangB2010},
corresponds to $L/d_{i}\to\infty$, therefore is not shown. \label{fig:Parameter-space}}

\end{figure}

\textcolor{black}{It is useful to map the numerical solutions discussed
below into a phase diagram, shown in Figure \ref{fig:Parameter-space}.
When the Hall effect is included, the system is characterized by two
important dimensionless parameters: $S$ and $L/d_{i}$. The parameter
space of $S$ and $L/d_{i}$ may be divided qualitatively into four
regions. The Hall reconnection regime is realized when the conventional
criterion $d_{i}>\delta_{SP}$ for onset of Hall reconnection is satisfied.
Under this condition, we recover the standard results for the onset
of Hall reconnection.\cite{MaB1996a,DorelliB2003,Bhattacharjee2004,CassakSD2005}
The Sweet-Parker reconnection regime is realized when neither the
criterion for onset of Hall reconnection $d_{i}>\delta_{SP}$ nor
that for the plasmoid instability $S>S_{c}$ are satisfied. In this
regime, a stable, elongated Sweet-Parker current layer is formed.
When the Lundquist number $S$ exceeds the critical value $S_{c}$
for onset of the plasmoid instability, two new possibilities emerge.
If the }secondary current sheets cascade down to widths at the $d_{i}$
scale, we may expect onset of Hall reconnection. On the other hand,
if the secondary current sheets never reach the $d_{i}$ scale, the
reconnection may proceed in a manner similar to that in resistive
MHD. To delineate\textcolor{black}{{} the border between these two regimes,
an estimate for the widths of secondary current sheets is needed.
In a previous resistive MHD study, we found that }a good estimate
for the average width of the secondary current sheets is obtained
by requiring that they obey Sweet-Parker scaling, with a length that
keeps them marginally stable. That gives an average width $\delta\sim\delta_{SP}(S_{c}/S)^{1/2}\sim LS_{c}^{1/2}/S$.\cite{HuangB2010}
We denote the regime where $\delta<d_{i}$ as {}``Hall reconnection
triggered by plasmoids'', and the regime where $\delta>d_{i}$ as
{}``plasmoid-dominated reconnection'' to characterize their different
possible behaviors. Note that statistical deviations from the average
width can and do occur.\cite{HuangB2010} As individual secondary
current sheets can be significantly thinner than the average width,
we expect the {}``Hall reconnection triggered by plasmoids'' region
to be larger than depicted in Figure \textcolor{black}{\ref{fig:Parameter-space}.}
We caution that since high-$S$, large-scale Hall MHD reconnection
is largely unexplored, Figure \textcolor{black}{\ref{fig:Parameter-space}}
cannot be regarded as a complete picture because it includes ranges
of parameter space where no simulations exist. Even the critical Lundquist
number $S_{c}$ and the secondary current sheet width $\delta$ could
be modified by the presence of the Hall effect.\textcolor{black}{{}
Furthermore, the criterion for onset $\delta<d_{i}$ is only accurate
up to a numerical factor of order unity. For these reasons, the delineation
of different regimes in Figure \ref{fig:Parameter-space} may not
be very precise. Nonetheless, }Figure \textcolor{black}{\ref{fig:Parameter-space}
serves well in guiding the choice of simulation parameters where interesting
physics may arise. }

\textcolor{black}{The primary interest of this work is to explore
the two new regimes where the plasmoid instability may play an important
role. This study includes three new runs (Run A to C), with corresponding
parameters marked on Figure \ref{fig:Parameter-space}. }A fourth
run, Run D, from a previous resistive MHD study\cite{HuangB2010},
is included for comparison. We fix $S=5\times10^{5}$ for all runs.
The parameter $L/d_{i}$ is $2.5\times10^{3}$ for Run A, $5\times10^{3}$
for Run B, and $10^{4}$ for Run C, respectively. We have chosen parameters
for the new runs such that after the onset of the plasmoid instability
the current sheets would have widths (estimated from the scaling law
$\delta\sim\delta_{SP}(S_{c}/S)^{1/2}$) ranging from $d_{i}$ (Run
A) to $4d_{i}$ (Run C). This is the parameter regime where we may
expect to observe a transition from the {}``Hall reconnection triggered
by plasmoids'' regime to the {}``plasmoid-dominated reconnection''
regime, depending on the ratio $\delta/d_{i}$. The initial condition
and governing parameters for these runs allow a clear separation of
length scales: the drivers of reconnection (the two merging islands)
are on the largest scale $\sim1$; the initial current layer width
$\sim0.01$; the Sweet-Parker width $\sim10^{-3}$; and the ion skin
depth $d_{i}\sim1-4\times10^{-4}$. Therefore, the simulations cover
all distinct stages of reconnection from the initial current sheet
thinning to the onset of plasmoid instability, which subsequently
may or may not lead to onset of Hall reconnection.

\section{Simulation Results}

\begin{figure}
\begin{centering}
\includegraphics[scale=0.6]{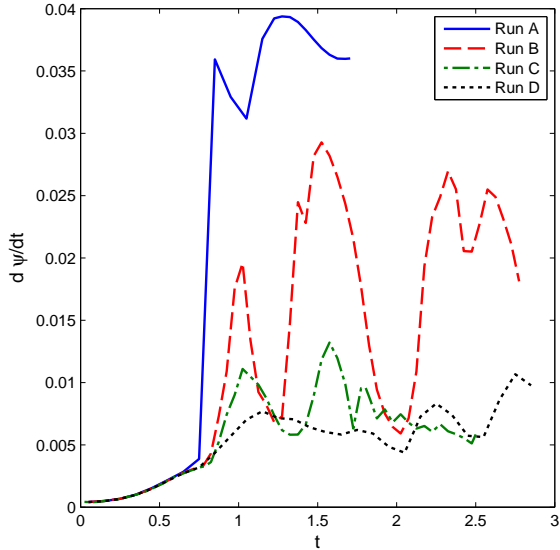}
\par\end{centering}

\caption{(Color online) The reconnection rate as a function of time, for four
different runs.\label{reconnection rate} }

\end{figure}

\begin{figure}
\begin{centering}
\includegraphics[scale=0.6]{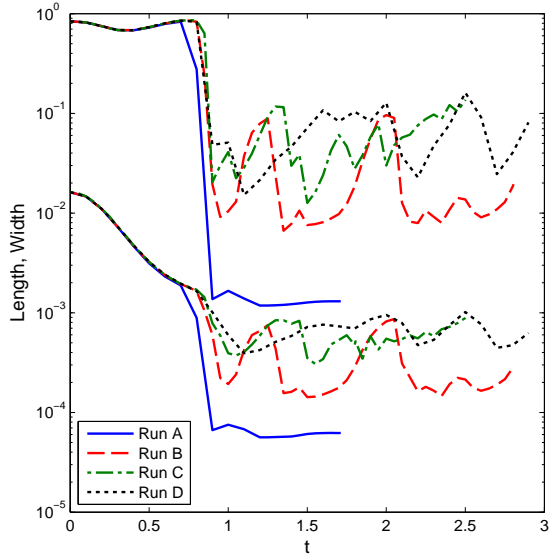}
\par\end{centering}

\caption{(Color online) The length (upper curve) and width (lower curve) of
the main reconnection current sheet as a function of time, for four
different runs. \label{fig:length-and-width}}

\end{figure}

\begin{figure}
\begin{centering}
\includegraphics[scale=0.45]{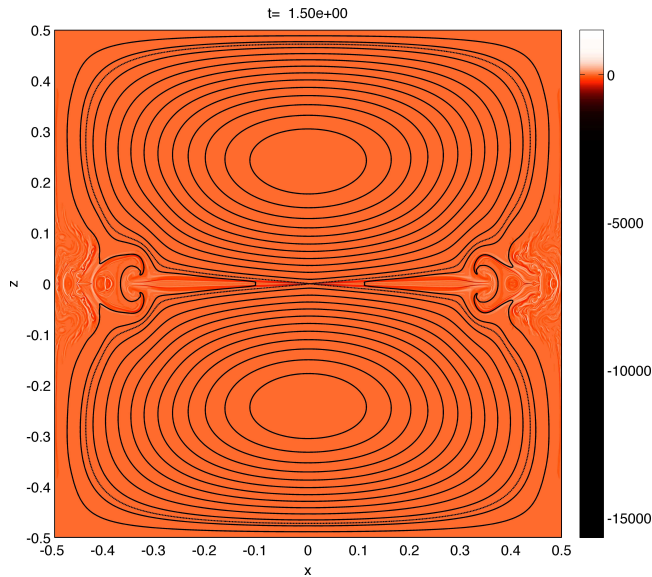}
\par\end{centering}

\caption{(Color online) Out-of-plane electric current density at $t=1.5$ for
Run A, overlaid with magnetic field lines, in the whole simulation
box. Dashed lines indicate separatrices, which are the field lines
that separate the two merging islands.\label{fig:RunA}}

\end{figure}

\begin{figure}
\begin{centering}
\includegraphics[scale=0.6]{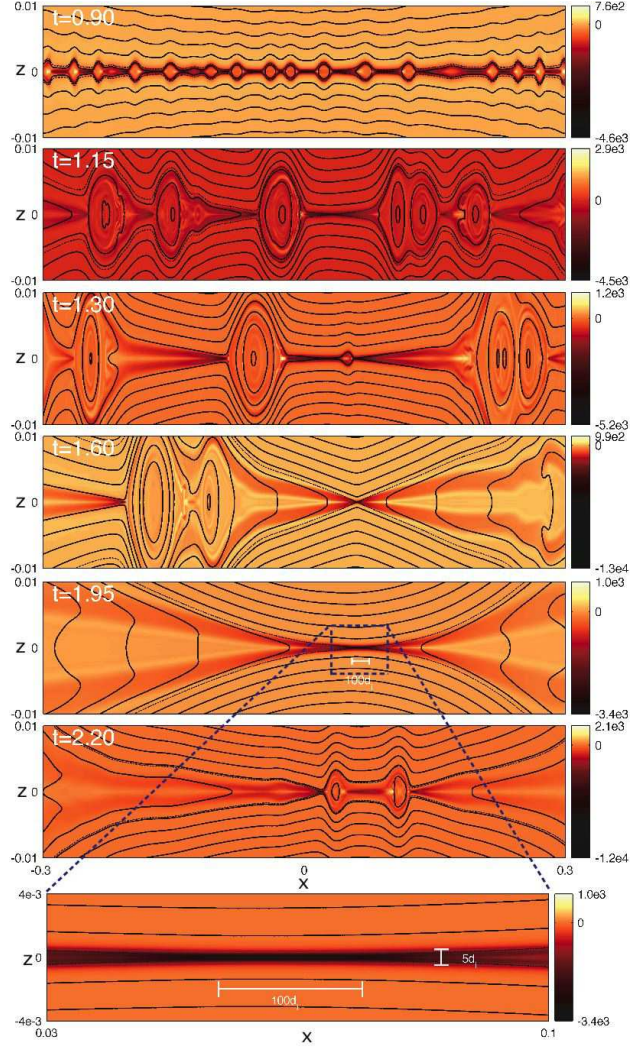}
\par\end{centering}

\caption{(Color online) Time sequence of the out-of-plane electric current
density for Run B, overlaid with magnetic field lines. Dashed lines
indicate separatrices. From top to bottom: (1) The Sweet-Parker current
sheet breaks up into a chain of plasmoids. (2) The plasmoids grow
in size; some of them are expelled to the downstream region; some
of them coalesce to form larger plasmoids. (3) A new plasmoid forms
at the main current sheet. (4) The formation of the new plasmoid leads
to an onset of Hall reconnection that eventually expels all plasmoids.
(5) The current sheet becomes extended again. (6) Subsequently, the
extended current sheet breaks up into plasmoids, which lead to another
onset of Hall reconnection. The bottom panel shows an expanded view
of the extended current sheet at $t=1.95$ (enhanced online). \label{fig:Time_seq}}
 
\end{figure}

\begin{figure}
\begin{centering}
\includegraphics[scale=0.6]{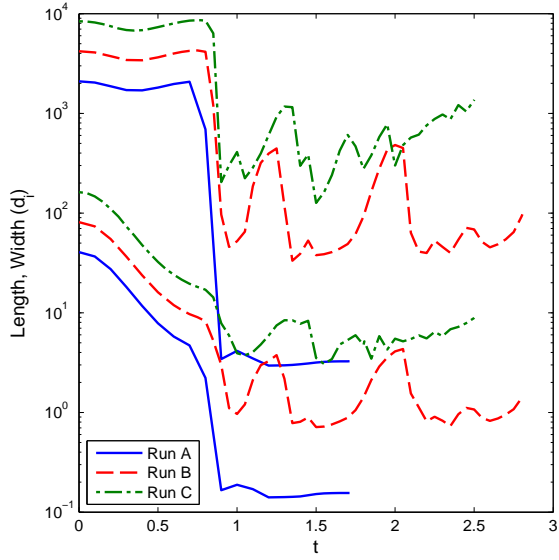}
\par\end{centering}

\caption{(Color online) The length (upper curve) and width (lower curve) of
the main reconnection current sheet, normalized to the ion skin depth
$d_{i}$, as a function of time for Run A to Run C. \label{fig:length-and-width-1}}

\end{figure}

\begin{figure}
\includegraphics{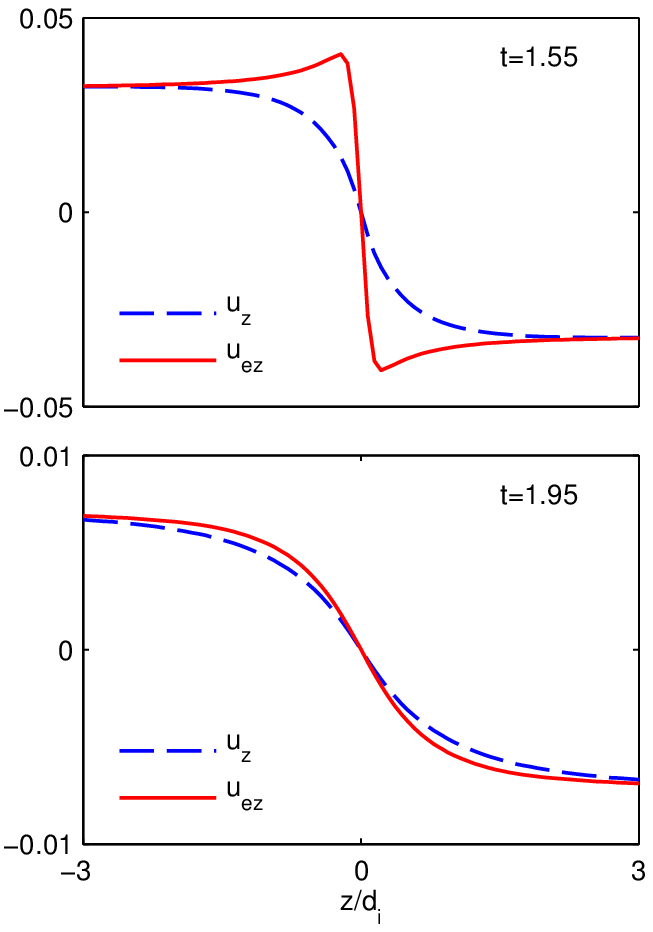}

\caption{(Color online) The electron and ion flows along the inflow ($z$)
direction through the X-point for Run B at $t=1.55$ and $t=1.95$.
\label{fig:vzcut}}

\end{figure}

Our primary diagnostics are the reconnection rate and the length and
width of the main reconnection current sheet. The reconnection rate
is measured as the time derivative of the reconnected magnetic flux.
In the presence of the plasmoid instability, the reconnection layer
generally contains multiple current sheets at a given time. We define
the main reconnection current sheet as the one located where the two
primary coalescing islands touch each other. This is the (generally
unique) point where the separatrix flux surface bounding the two merging
islands intersects itself. For example, in the second panel of Figure
\ref{fig:Time_seq}, the main reconnection current sheet is the one
at the center, between $x=0$ and $x=0.1$. The length and width are
measured as the full width at quarter maximum.

Figure \ref{reconnection rate} and \ref{fig:length-and-width} show,
respectively, the time-histories of the reconnection rate, and the
length and width of the main reconnection current sheet for four different
runs. Initial current sheet thinning occurs from $t=0$ to $t=0.7$.
During this period, the four runs are very similar because the Hall
current has yet to play an important role. The current sheet width
thins from the initial $\delta\sim10^{-2}$ down to $\delta\sim\delta_{SP}\sim10^{-3}$.
Meanwhile, the reconnection rate gradually rises to $3\times10^{-3}$.
The plasmoid instability sets in at approximately $t=0.7$. Thereafter,
the three new runs exhibit qualitatively different behaviors. In Run
A, the plasmoid instability immediately triggers a strong onset of
Hall reconnection, which expels all the plasmoids, and the system
is left with a single X-point. After that, the system reaches a quasi-steady
state with the reconnection rate and current sheet geometry approximately
time-independent. This run gives the highest reconnection rate (up
to $0.04$) of the four runs, and the current sheet is also the shortest
and narrowest. The aspect ratio (width/length) of the current sheet
in the quasi-steady state is approximately $1/20$. Figure \ref{fig:RunA}
shows the out-of-plane electric current density, overlaid with magnetic
field lines, in the whole domain at $t=1.5$. Dashed lines denote
the separatrices which are the field lines that separate the two merging
islands. The reconnection site clearly shows a Petschek-like geometry
with the separatrices opening up in the downstream region.

In Run B (see Figure \ref{fig:Time_seq} for a few snapshots of the
key stages), the plasmoid instability does not immediately lead to
onset of Hall reconnection. An onset occurs at approximately $t=1.3$,
triggered by a new plasmoid formed in the main reconnection current
sheet. Subsequently all plasmoids are wiped out. However, it appears
that Hall reconnection with a single X-point is unstable for this
set of parameters, and the system makes a transition back to an extended
current sheet. The current sheet length reaches a maximum $(\simeq0.1=500d_{i})$
at $t=2$, whereupon it becomes unstable again and breaks up into
plasmoids. This second onset of plasmoid instability leads to another
onset of Hall reconnection, resulting in a single X-point again. Conceivably,
this cycle will continue repeatedly until the system runs out of flux.
Indeed, towards the end of this run, we observe that the length and
width of the main current sheet start to rise again (Figure \ref{fig:length-and-width}).
In this regime, which we have called the intermediate regime, the
system is caught in between Hall reconnection with a single X-point,
and plasmoid-dominated reconnection with multiple X-points. The resulting
reconnection rate fluctuates strongly between $0.005$ to $0.03$.
For Run C, because $d_{i}$ is well below the smallest scale caused
by the plasmoid instability, the system never makes a transition to
Hall reconnection. The reconnection rate from Run C, ranging from
$0.005$ to $0.013$, is similar to that from Run D, which is a resistive
MHD simulation ($d_{i}=0$). 

In Figure \ref{fig:length-and-width-1} we replot the time histories
of the length and width of the main current sheet as shown in Figure
\ref{fig:length-and-width}, but this time in units of $d_{i}$. Hall
reconnection is characterized by the decoupling of ions and electrons
at scales smaller than $d_{i}$, and the dissipation region where
the frozen-in condition is broken is significantly smaller than the
$d_{i}$ scale. Run A clearly exhibits these features, as the current
sheet width during the quasi-steady phase is approximately $0.15d_{i}$.
On the other hand, the main current sheet in Run B is never significantly
thinner than $d_{i}$. The minimum current sheet width is approximately
$0.7d_{i}$ in this run. (However, note that we measure the current
sheet width by its full width at quarter maximum. Instead, if we measure
by its half width at half maximum, as employed by Cassak \emph{et
al.}\cite{CassakSD2005}, the minimum width in Run B is $0.17d_{i}$.
This value is on par with the typical current sheet width of Hall
solutions reported by Cassak \emph{et al.}\cite{CassakSD2005}) This
suggests that the Hall reconnection after onset is not as robust as
it is in Run A. Nonetheless, Run B clearly shows the characteristic
of Hall reconnection, \emph{i.e.} the decoupling of electron and ion
flow at scales below $d_{i}$, when the current sheet width reaches
the minimum. Figure \ref{fig:vzcut} shows one-dimensional (1D) profiles
of the electron and ion flows along the $z$ direction through the
X-point, at $t=1.55$ and $1.95$. At $t=1.55$, the electron and
ion flows are clearly decoupled, indicating that the reconnection
is in the Hall regime. On the other hand, when the current sheet becomes
elongated again at $t=1.95$, the electron and ion flows closely follow
each other, indicating that Hall current does not play an important
role at this time. The current sheet width in Run C only reaches a
minimum of approximately $3d_{i}$, which is why Run C never shows
any onset of Hall reconnection.

\section{Discussion\label{sec:Discussion}}

\begin{figure}
\begin{centering}
\includegraphics[scale=0.6]{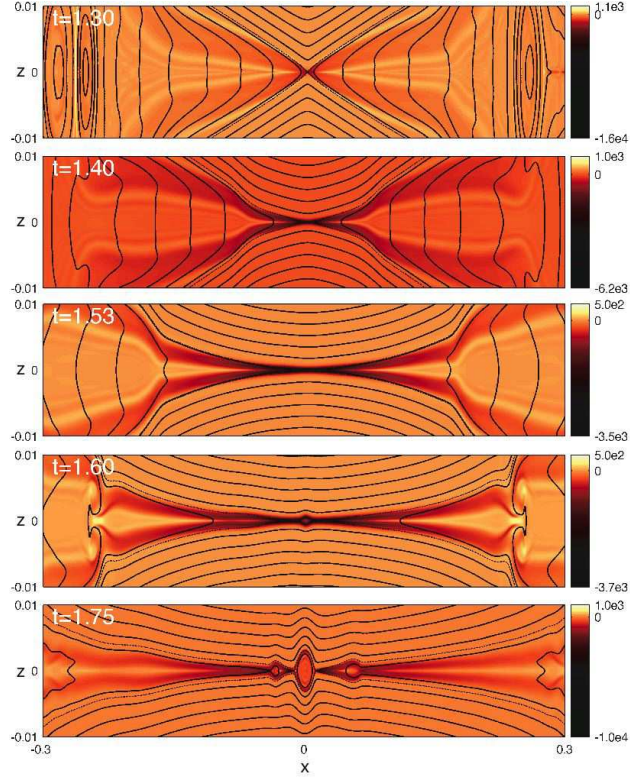}
\par\end{centering}

\caption{(Color online) Time sequence of the out-of-plane electric current
density for the artificial test, overlaid with magnetic field lines.
The initial condition is taken from Run A at $t=1.3$, with the ion
skin depth $d_{i}$ artificially lowered from $4\times10^{-4}$ to
$2\times10^{-4}$, which is the same as Run B. The opening angle between
the separatrices quickly closes up, first starting from the center,
then gradually propagating outward. As the current sheet becomes extended,
it becomes unstable to the plasmoid instability (enhanced online).\label{fig:artificial}}

\end{figure}

An important question is, why does Run B revert to an extended current
sheet after the onset of Hall reconnection? To answer this question,
it is important to appreciate that although the global Lundquist number
$S$ is high ($S=5\times10^{5}$) for these runs, the resistivity
is not negligible on the length scale of $d_{i}$. This is because
$L/d_{i}$ is also a large number, which is often the case in many
plasmas of interest. A relevant dimensionless parameter that quantifies
how resistive the plasma is on the $d_{i}$ scale is the Lundquist
number based on $d_{i}$, defined as $S_{d_{i}}\equiv V_{A}d_{i}/\eta$.
For Run A, B, and C, the Lundquist numbers based on $d_{i}$ are $200$,
$100$, and $50$, respectively. 

Recently, it has been demonstrated by Cassak \emph{et al}. that over
a certain range of $S_{d_{i}}$, resistive Hall reconnection exhibits
bistability, \emph{i.e.} both Sweet-Parker and Hall reconnection are
physically realizable.\cite{CassakSD2005,CassakSD2010}(Cassak \emph{et
al}. use the notation $\eta'$, which is $1/S_{d_{i}}$.) The condition
for bistability may be expressed as\cite{CassakSD2005} \begin{equation}
\frac{L}{d_{i}}>S_{d_{i}}>S_{d_{i}}^{c},\label{eq:bistability}\end{equation}
where $S_{d_{i}}^{c}$ is a critical Lundquist number based on the
$d_{i}$ scale. Here the condition $L/d_{i}>S_{d_{i}}$ is equivalent
to the condition $\delta_{SP}>d_{i}$ for the existence of the Sweet-Parker
solution, and the condition $S_{d_{i}}>S_{d_{i}}^{c}$ simply means
that the plasma cannot be too resistive on the $d_{i}$ scale, otherwise
the Hall solution will cease to exist. If $S_{d_{i}}>L/d_{i}$, only
the Hall solution is available; and if $S_{d_{i}}<S_{d_{i}}^{c}$,
only the Sweet-Parker solution is realizable. The critical value $S_{d_{i}}^{c}$
was found to be approximately $50$ in Ref. \cite{CassakSD2005} for
a double tearing mode configuration with two Harris current sheets
in a system of dimensions $409.6d_{i}\times204.8d_{i}$. That study
also included electron inertia, with a mass ratio of $m_{e}/m_{i}=1/25$. 

In the present study, it appears that Run B, with $S_{d_{i}}=100$,
is already below the critical value $S_{d_{i}}^{c}$ for transition,
therefore the Hall solution is unstable. This implies that the critical
value $S_{d_{i}}^{c}$ is greater than $100$, higher than approximately
$50$ found by Cassak \emph{et al.} To verify that the Hall solution
is unstable for the set of parameters of Run B, we carry out the following
test. We take the solution of Run A at $t=1.3$ and restart with the
ion skin depth $d_{i}$ artificially lowered to $2\times10^{-4}$,
the same value as Run B. Figure \ref{fig:artificial} shows the time
sequence of this test. As a result of lowering $d_{i}$, the opening
angle between the separatrices quickly closes up, first starting from
the center, then gradually propagating outward. The current sheet
becomes extended at the same time and eventually breaks up into plasmoids.
This test confirms that the Hall solution is indeed unsustainable
for the set of parameters of Run B. On the other hand, the other solution
that the system will tend to make a transition to --- the Sweet-Parker
solution --- is also unstable due to the plasmoid instability. Therefore,
Run B is in a {}``bi-unstable'' regime, and shows a continuous generation
of new plasmoids.

From the above discussion, it is now clear that to realize the intermediate
regime, we need the following two conditions: First, secondary current
sheets must be able to reach the $d_{i}$ scale to trigger Hall reconnection.
Second, the Hall solution has to be unstable, \emph{i.e.} the condition
$S_{d_{i}}<S_{d_{i}}^{c}$ must be satisfied. Therefore, to delineate
the region of the intermediate regime in the parameter space, it is
important to know how the critical value $S_{d_{i}}^{c}$ depends
other dimensionless parameters. Cassak \emph{et al.} \cite{CassakSD2005}
give an estimate of $S_{d_{i}}^{c}$ by equating the resistive diffusion
across the current sheet $\eta/\delta^{2}$ with the inward convection
$u_{in}/\delta$, where $\delta$ is the current sheet width and $u_{in}$
is the electron inflow speed. They assume that the current sheet width
$\delta$ scales like the electron skin depth $d_{e}=c/\omega_{pe}$,
where $\omega_{pe}$ is the electron plasma frequency, and $u_{in}$
scales like $0.1V_{Ae}$, where $V_{Ae}=B/\sqrt{4\pi nm_{e}}$ is
the electron Alfv\'en  speed based on the magnetic field immediately
upstream of the electron current layer. By using the observed upstream
magnetic field $B\sim0.3B_{0}$, where $B_{0}$ is the asymptotic
field, they obtain an estimate $S_{d_{i}}^{c}\sim30$, which is reasonably
close to the observed value $S_{d_{i}}^{c}\simeq50$. 

The above argument might suggest that $S_{d_{i}}^{c}$ is a constant
independent of system parameters, but this is open to question. For
example, $S_{d_{i}}^{c}$ may depend on the ratios $L/d_{i}$ and
$m_{e}/m_{i}$. This issue cannot be settled by appealing to the numerical
data presented in Ref. \cite{CassakSD2005}, which presents results
for only one set of ratios ($L/d_{i}=409.6$ and $m_{e}/m_{i}=1/25$).
In the present study, electron inertia is neglected. Therefore, we
focus on the possible dependency of $S_{d_{i}}^{c}$ on $L/d_{i}$.
To determine $S_{d_{i}}^{c}$ for each system size requires many runs
and large computational resources, which are not practical at the
present time. Instead, we perform a series of simulations with $S_{d_{i}}=100$,
same as Run B, but with smaller system sizes. Our results suggest
that $S_{d_{i}}^{c}$ increases with increasing system size $L/d_{i}$.

We perform two simulations with $L/d_{i}=500,$ $1000$ and $S=5\times10^{4}$,
$10^{5}$, respectively. Also, we restart Run A at $t=1.3$, but lower
the Lundquist number $S$ to $2.5\times10^{5}$. As such, we have
three additional runs with $L/d_{i}=500,$ $1000$, and $2500$, all
with the same $S_{d_{i}}=100$. For the first two runs, the Sweet-Parker
layer becomes unstable to the plasmoid instability, which quickly
triggers onset of Hall reconnection. The onset of Hall reconnection
expels all the plasmoids, and the reconnection precedes in the same
manner as Run A. For the third run, the diffusion region broadens
quickly after the Lundquist number is lowered, with the current sheet
width increasing from $0.15d_{i}$ to $0.65d_{i}$. After that, the
system remains in a quasi-steady X-point geometry. These results indicate
that for these smaller system sizes, the critical value $S_{d_{i}}^{c}$
is smaller than $100$; whereas for Run B, with $L/d_{i}=5000$, the
critical value $S_{d_{i}}^{c}$ is greater than $100$. 

\begin{figure}
\includegraphics{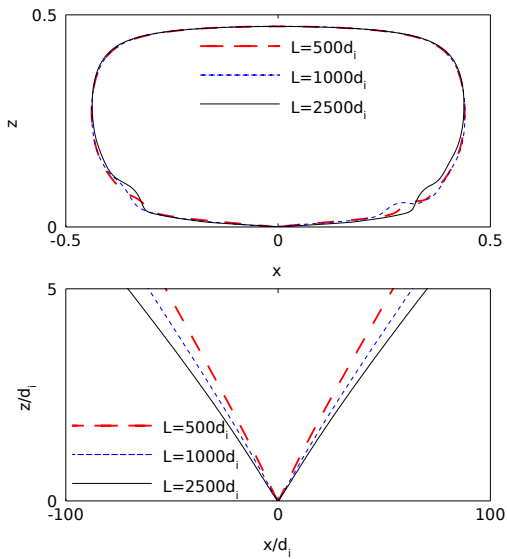}

\caption{(Color online) The separatrices of the three runs when the reconnected
fluxes are approximately the same ($\simeq0.027$). Only the region
$z>0$ is shown. Upper panel: the whole simulation domain. Lower panel:
a close-up view around the X-point. Note that in the lower panel the
coordinates are normalized to $d_{i}$ and shifted horizontally to
account for the slight misalignment of the X-point for each run. Also
the $z$ direction is stretched for better visualization.\label{fig:separatrix}}

\end{figure}

\begin{figure}
\includegraphics{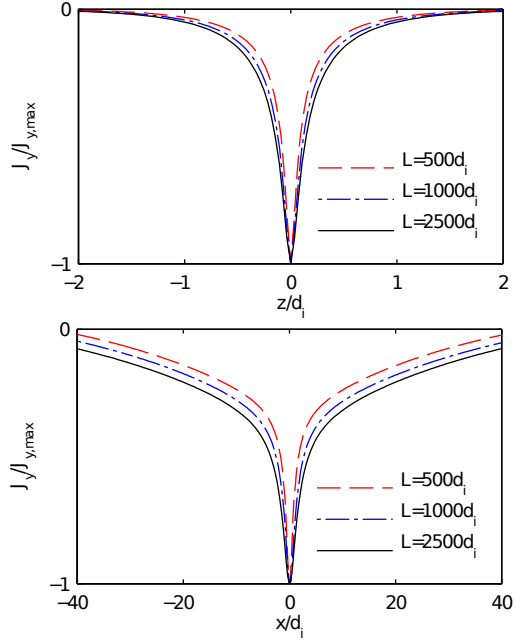}

\caption{(Color online) Current density profiles alone the inflow (upper panel)
and the outflow (lower panel) directions. Here we normalize the coordinates
to $d_{i}$ and the current density to the peak value. \label{fig:Jycut}}

\end{figure}

\begin{figure}
\includegraphics{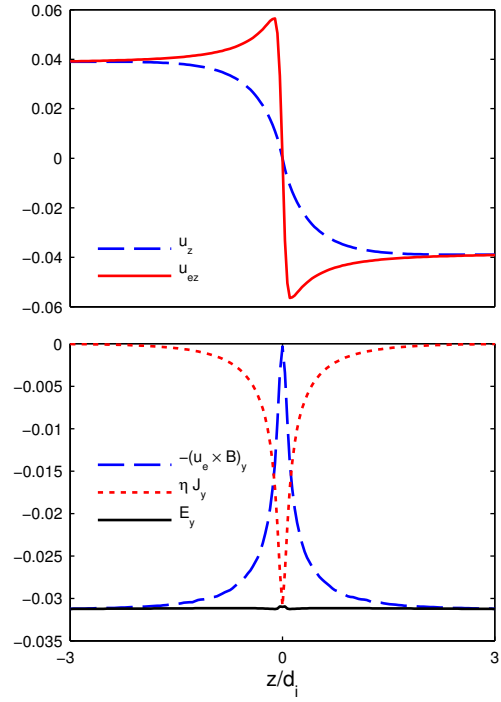}

\caption{(Color online) The electron and ion inflows (upper panel) and the
balance of $E_{y}=-(\mathbf{u}_{e}\times\mathbf{B})_{y}+\eta J_{y}$
(lower panel) in the generalized Ohm's law, for the case $L/d_{i}=1000$\label{fig:ebalance}}

\end{figure}

A closer comparison of the dissipation regions for these runs reveals
some interesting features. We observe a consistent trend that in the
quasi-steady phase, the larger the system size is, the smaller is
the opening angle in the downstream region. Furthermore, the current
sheet is found to be longer and wider, when normalized to $d_{i}$,
for a larger system. Figure \ref{fig:separatrix} shows the separatrices
of the three runs when the reconnected fluxes are approximately the
same ($\simeq0.027$). The upper panel shows the entire simulation
domain. The proximity of the three curves indicates that the global
conditions are similar. However, a close-up view around the X-point,
shown in the lower panel, reveals that the opening angle in the downstream
region is smaller for a larger system.  Likewise, in Figure \ref{fig:Jycut}
we plot current density profiles along the inflow and the outflow
directions. Here we normalize lengths to $d_{i}$ and the current
density to the peak value. Clearly, the current sheet length and width
increase with increasing the system size $L/d_{i}$. Finally, these
runs are all well resolved with more than $30$ grid points per $d_{i}$
along the inflow direction at the current sheet. Figure \ref{fig:ebalance}
shows the electron and ion inflows and the balance of $E_{y}=-(\mathbf{u}_{e}\times\mathbf{B})_{y}+\eta J_{y}$
in the generalized Ohm's law, for the case $L/d_{i}=1000$. The $-(\mathbf{u}_{e}\times\mathbf{B})_{y}$
term and $\eta J_{y}$ term add up to a nearly uniform out-of-plane
electric field $E_{y}$, as required for quasi-steady reconnection
in two dimensions. This indicates that the current sheet is well resolved
and the reconnection is supported by resistivity, rather than numerical
dissipation.

The fact that the current sheet width $\delta$ increases monotonically
with increasing $L/d_{i}$ for the same $S_{d_{i}}$ suggests that
the critical value $S_{d_{i}}^{c}$ also increases monotonically with
$L/d_{i}$. This is evident from the following thought experiment.
Suppose we start from a Hall solution, and gradually lower $S_{d_{i}}$
by increasing $\eta$, the current sheet width $\delta$ will gradually
increase. Because the current sheet in Hall reconnection has to be
thinner than $d_{i}$, the Hall solution will cease to exist when
the current sheet width is approaching $O(d_{i})$. (More precisely,
the Hall solution will cease to exist when its width is equal to the
width of the unstable solution\cite{CassakDSE2007},\emph{ i.e.} when
the stable fixed point and the unstable fixed point annihilate each
other; see the discussion in Ref. \cite{CassakSD2010}.) Since the
current sheet width increases in a larger system for the same $S_{d_{i}}$,
the width will approach $O(d_{i})$ at a higher $S_{d_{i}}$ in a
larger system. Consequently, a larger system has a higher critical
value $S_{d_{i}}^{c}$. This conclusion is consistent with our finding
that $S_{d_{i}}^{c}>100$ for $L/d_{i}=5000$ while $S_{d_{i}}^{c}<100$
for the three smaller systems.

Intuitively, the dependence of the dissipation region on the system
size may be understood as a competition between the attraction of
the two coalescing islands on the global scale, which tends to close
up the downstream region and make the current sheet extended, and
the Hall physics at the local $d_{i}$ scale, which opens up the downstream
region. Our results indicate that to have a complete understanding
of what determines $S_{d_{i}}^{c}$, we need a theory that couples
local reconnection physics with the global configuration. Although
the precise scaling is unknown at this time, the fact that $S_{d_{i}}^{c}$
increases with increasing $L/d_{i}$ has profound implications on
the accessibility of the intermediate regime in large systems. Note
that the line that separates region (3) and region (4) in Figure \ref{fig:Parameter-space}
has a constant $S_{d_{i}}$. Therefore, if $S_{d_{i}}^{c}$ is independent
of $L/d_{i}$, conceivably the intermediate regime will be a narrow
region between region (3) and region (4). Now we have shown that $S_{d_{i}}^{c}$
increases with increasing $L/d_{i}$. That implies that it will be
easier to access the intermediate regime for larger systems. Therefore,
it is important to determine the precise scaling of $S_{d_{i}}^{c}$
with respect to $L/d_{i}$. Such a study requires the investment of
significant computational resources, and is left to the future work.
Finally, when more sophisticated models are employed, the condition
for transition may also depend on some other dimensionless parameters
as well.

The existence of the intermediate regime and the dependence of $S_{d_{i}}^{c}$
on $L/d_{i}$ in other global configurations remain to be studied.
The present island coalescence configuration differs from the more
commonly studied tearing mode configuration in one important aspect.
The island coalescence configuration has the attractive force between
the two merging islands as an {}``ideal'' drive of reconnection,
which is absent in the tearing mode configuration. Could this be the
reason that the intermediate regime has not been found yet in the
standard tearing mode configuration? We do not know the answer. However,
it should be noted that previous Hall MHD simulations all have substantially
smaller system sizes, which, according to our findings, may make the
intermediate regime less likely to occur. Only future studies with
much larger system sizes could possibly answer this question. 

In light of the present study, we would like to comment on the recent
controversy regarding the role of electron inertia effects on bistability.\cite{ZoccoCS2009,CassakSD2010,SullivanBH2010}
Whereas Zocco \emph{et al.} \cite{ZoccoCS2009} claim that electron
inertia is essential, Cassak \emph{et al. }\cite{CassakSD2010} argue
that it is Hall physics rather than electron inertia that is responsible
for bistability. The conclusion of Cassak \emph{et al.} is supported
by an independent study by Sullivan \emph{et al.} \cite{SullivanBH2010}
Run A in the present study may be interpreted as an independent verification
of the claim made by Cassak \emph{et al.} \cite{CassakSD2010} and
Sullivan \emph{et al.} \cite{SullivanBH2010} that bistability survives
even in the absence of electron inertia. In Run A, the Hall solution
is realized and remains stable after onset of the plasmoid instability.
However, the Sweet-Parker solution clearly exists, because the Sweet-Parker
width $\delta_{SP}\simeq10^{-3}$ is significantly above the ion skin
depth $d_{i}=4\times10^{-4}$. Had it not been for the intervention
of the plasmoid instability, Run A would have realized the Sweet-Parker
solution. Therefore, within the framework of the original bistability
theory, when the plasmoid instability is not taken into account, both
Sweet-Parker and Hall solutions are realizable for the set of parameters
of Run A, and the system is bistable. When the plasmoid instability
is included, the Sweet-Parker solution in Run A becomes physically
unrealizable, and the Hall solution is the only possibility. The present
study is valuable as an independent test of the phenomenon of bistability
because it is done with a different code and a different initial condition.
(Both Cassak \emph{et al. }and Sullivan \emph{et al.} use the code
F3D.\cite{ShayDSR2004}) 

However, Cassak \emph{et al.} \cite{CassakSD2010} further claim that
balancing the Hall term and the resistive term in the out-of-plane
component of the generalized Ohm\textquoteright{}s law (as was done
in Refs. \cite{ChaconSZ2007,SimakovC2008,ZoccoCS2009,Malyshkin2008,Malyshkin2009})
corresponds to an unstable, and thus physically unrealizable, solution.
\textcolor{black}{As such, compressing the current layer leads to
a runaway toward smaller scales. They argue that the runaway process
{}``stops only when additional physics, such as off-diagonal elements
of the pressure tensor, become important at electron scales,'' and
{}``in two-fluid simulations of Hall reconnection, the runaway process
is often stopped using an explicit high order dissipation term such
as hyperviscosity or through numerical dissipation because off-diagonal
pressure tensor terms are absent from the model.''}\textcolor{blue}{{}
}(See the discussion in Sec. II of Ref. \cite{CassakSD2010}.)\textcolor{blue}{{}
}\textcolor{black}{This claim is inconsistent with our own study,
where we have found that the $\eta J_{y}$ term by itself can balance
the reconnection out-of-plane electric field around the X-point in
robust Hall reconnection without the assistance of higher order dissipation
terms, as shown in Figure \ref{fig:ebalance}. Supporting the reconnection
electric field solely by resistivity has been shown before by Wang
}\textcolor{black}{\emph{et al. }}\textcolor{black}{\cite{WangBM2001}
and recently by }Sullivan \emph{et al.} \cite{SullivanBH2010}. Along
with this study, we have carefully verified this result via convergence
tests on smaller systems. 

Because resistivity by itself can stop the current sheet from collapsing
in a Hall solution without the need for other physics on electron
scales, the assumption made in the argument by Cassak \emph{et al}.
\cite{CassakSD2005} that the current sheet width in Hall solution
scales as $d_{e}$ is debatable in a resistive plasma. For example,
the current sheet in Figure \textcolor{black}{\ref{fig:ebalance}
is significantly wider than $d_{e}\simeq d_{i}/43$, when the real
mass ratio of a hydrogen plasma is employed. An estimate of the contribution
from the neglected electron inertia terms in the generalized Ohm's
law indicates that those terms are much smaller, therefore the assumption
of neglecting them is justified. }

We emphasize that although resistivity can in principle balance the
reconnection electric field in a Hall solution, such a balance need
not necessarily be the case in Nature. It is clear that if the current
sheet width approaches the $d_{e}$ scale, electron physics will come
into play. However, in the thought experiment discussed above, the
current sheet width increases as we increase $\eta$. There exists
a certain range of $\eta$ where the current sheet width is above
$d_{e}$ but the solution remains in the Hall branch. In this range
of $\eta$ the reconnection electric field should mostly be balanced
by the $\eta J$ term. This thought experiment has been carried out
numerically by Cassak \emph{et al. }\cite{CassakSD2005} Contrary
to what we have suggested, the simulations show that the current sheet
widths in the Hall branch remain close to the $d_{e}$ scale when
$\eta$ increases, before a sudden transition to a much broader current
sheet in the Sweet-Parker branch (see Figure 3 of Ref. \cite{CassakSD2005}).
This may be due to the high electron mass ($m_{e}=m_{i}/25$) employed,
and consequently the $d_{e}$ and $d_{i}$ scales are not sufficiently
well separated. 

In summary, we find that the bistability theory of Cassak \emph{et
al.} remains a very useful concept even in the presence of the plasmoid
instability, and it greatly helps in interpreting our simulation results.
However, our results also indicate the present understanding of what
governs the extension of the current sheet and the stability of a
Hall solution is still incomplete. In particular, much needs to be
done regarding how global conditions may affect the local reconnection
site.\cite{KuritsynJGRY2007}

\section{Conclusion}

Our results show that the transition to fast reconnection in large,
high-Lundquist-number plasmas can be realized by a complex interplay
between the plasmoid instability and Hall reconnection. We have clearly
demonstrated that the plasmoid instability can facilitate the onset
of Hall reconnection, in a regime where Hall reconnection would otherwise
remain inaccessible because the criterion $d_{i}>\delta_{SP}$ is
not met (Runs A and B). However, the onset of Hall reconnection does
not always lead to a single X-point topology, with all plasmoids expelled.
Run B demonstrates the possibility that a single X-point geometry
is itself unstable, and after the onset of Hall reconnection, reverts
to an extended current sheet of the type that led to an X-point in
the first place. In this case, the reconnection is characterized by
sporadic, bursty behavior with new plasmoids constantly being generated.
Because of the intermittent onset of Hall reconnection, on average
the reconnection rate is higher than it is when the plasmoid instability
does not trigger Hall reconnection (Run C), but lower than it is when
a robust Hall reconnection site forms (Run A). 

The results presented here may provide a possible starting point to
resolve a dichotomy in the existing literature --- the X-point geometry
in Hall reconnection\textcolor{black}{\cite{MaB1996a,DorelliB2003,CassakSD2005,ShepherdC2010},
versus the extended current sheet geometry embedded with plasmoids
in fully kinetic simulations}\cite{DrakeSCS2006,DaughtonSK2006,DaughtonRAKYB2009,DaughtonRAKYB2009a,KlimasHZK2010}\textcolor{black}{.
Our results demonstrate that the dichotomy is false. We have shown
that for some range of parameters (Run B) resistive Hall MHD allows
the current sheet to become extended again after the onset, and subsequently
new plasmoids are generated. That is not to say that the physical
mechanisms that cause the extension of the current sheet is the same
in the present Hall MHD simulations and fully kinetic simulations.
Full kinetic simulations show extended current sheet and plasmoid
formation even in the absence collisions, }\cite{DrakeSCS2006,DaughtonSK2006,KlimasHZK2010}
which is not possible in the present simple fluid model, as lack of
collisions means $\eta\to0$ in the present model. Even when Run B
is compared with collisional PIC simulations \cite{DaughtonRAKYB2009,DaughtonRAKYB2009a}
there are discernible differences. For example, PIC simulations show
a continuous generation of new plasmoids. That is quite different
from Run B, where the reconnection geometry goes to a singe X-point
configuration first then becomes extended again, which triggers plasmoid
formation. 

The results of this work may be tested in the next generation of Magnetic
Reconnection Experiment (MRX), which is planned to systematically
explore different regimes in the reconnection {}``phase diagram''.
\cite{JiYPDR2010} It may also be applicable to magnetic reconnection
in laser produced high energy density plasmas, which have been the
subject of great interest recently.\cite{Nilson2006,Li2007,Nilson2008,Willingale2010,Zhong2010,Fox2011}
For applications in space and astrophysical systems, it is clear that
if Spitzer resistivity is assumed, the intermediate regime in the
present study is unlikely to be relevant in systems such as solar
corona and Earth's magnetosphere, where a simple estimate gives $S_{d_{i}}\sim10^{7}$
for solar corona (assuming $n\sim10^{9}cm^{-3},$ $B\sim100G$, $T\sim100eV$)
and $S_{d_{i}}\sim10^{10}$ for Earth's magnetotail (assuming $n\sim1cm^{-3}$,
$B\sim10^{-4}G$, $T\sim100eV$). \textcolor{black}{It should be borne
in mind, however, that the applicability of Spitzer resistivity in
those systems is also open to debate, as the relevant resistivity
may be due to wave particle interaction or other mechanisms.} On the
other hand, in solar chromosphere, due to variation in the plasma
density of about seven orders of magnitude, $S_{d_{i}}$ based on
Spitzer resistivity varies from $10^{-3}$ to $10^{4}$ (assuming
$n\sim10^{10}-10^{17}cm^{-3}$, $B\sim100G$, $T\sim1eV$ ). Therefore,
it is likely that there is some region in the chromosphere where the
intermediate regime is directly applicable. 

In conclusion, although the resistive Hall MHD model has limitations,
the fact that the single X-point geometry is not inevitable in the
Hall MHD model opens the possibility of realizing extended current
sheets in global Hall MHD simulations of large systems. In future
work, through the implementation of more sophisticated closures, \emph{e.g.}
for the electron pressure tensor term in the generalized Ohm\textquoteright{}s
law, one may be able to parameterize some key kinetic effects in reconnection
simulations.\textcolor{black}{{} Progress along this direction may be
essential in order to extend global modeling codes to include two-fluid
and kinetic effects, as fully kinetic simulations of large systems,
with realistic physical parameters are likely to remain computationally
too expensive even in the near-future era of exascale computing. }
\begin{acknowledgments}
\textcolor{black}{This work is supported by the Department of Energy,
Grant No. DE-FG02-07ER46372, under the auspice of the Center for Integrated
Computation and Analysis of Reconnection and Turbulence (CICART),
the National Science Foundation, Grant No. PHY-0215581 (PFC: Center
for Magnetic Self-Organization in Laboratory and Astrophysical Plasmas),
NASA Grant Nos. NNX09AJ86G and NNX10AC04G, and NSF Grant Nos. ATM-0802727,
ATM-090315 and AGS-0962698. YMH is partially supported by a NASA subcontract
to the Smithsonian Astrophysical Observatory's Center of Astrophysics,
Grant No. NNM07AA02C.  Computations were performed on facilities at
National Energy Research Scientific Computing Center. YMH would like
to thank Dr. Naoki Bessho for commenting on an earlier version of
the paper, and Prof. Paul Cassak for beneficial communications when
the paper was under revision. We are grateful for the comments of
the anonymous reviewers, which have helped improve this work.}\end{acknowledgments}

\end{document}